\documentclass[aps,pra,reprint,groupedaddress]{revtex4-1}
\usepackage{graphicx}
\usepackage{color}
\usepackage{bm}
\usepackage{comment}
\usepackage{amssymb, amsmath}

\begin{document}


\title{Quantum Gate for Kerr Nonlinear Parametric Oscillator Using Effective Excited States}

\author{Taro Kanao$^1$}
\email[]{taro.kanao@toshiba.co.jp}
\author{Shumpei Masuda$^2$}
\author{Shiro Kawabata$^2$}
\author{Hayato Goto$^1$}
\affiliation{$^1$Frontier Research Laboratory, Corporate Research \& Development Center, Toshiba Corporation, 1, Komukai-Toshiba-cho, Saiwai-ku, Kawasaki 212-8582, Japan\\
	$^2$Research Center for Emerging Computing Technologies (RCECT), National Institute of Advanced Industrial Science and Technology (AIST), 1‑1‑1, Umezono, Tsukuba, Ibaraki 305‑8568, Japan}


\date{\today}

\begin{abstract}
A Kerr nonlinear parametric oscillator (KPO) can stabilize a quantum superposition of two coherent states with opposite phases, which can be used as a qubit. In a universal gate set for quantum computation with KPOs, an $R_x$ gate, which interchanges the two coherent states, is relatively hard to perform owing to the stability of the two states. We propose a method for a high-fidelity $R_x$ gate by exciting the KPO outside the qubit space with parity-selective transitions, which can be implemented by only adding a driving field. In this method, the utilization of higher effective excited states leads to a faster $R_x$ gate, rather than states near the qubit space. The proposed method can realize a continuous $R_x$ gate and thus is expected to be useful for, e.g., recently proposed variational quantum algorithms.
\end{abstract}


\maketitle

\section{Introduction}\label{sec_intro}
Quantum computation using two coherent states as a qubit has been proposed in an optical circuit~\cite{Ralph2003} and a superposition of two coherent states has been an important resource state.
Recently, the generation of such a superposition has been designed~\cite{Mirrahimi2014} and realized experimentally~\cite{Leghtas2015} in a superconducting circuit by engineering a two-photon drive and two-photon loss.
Also by this approach, a coherent state has experimentally been stabilized, where a dominant error source is single-photon loss~\cite{Lescanne2020}.

In a superconducting circuit, another method to stabilize a superposition of coherent states has been proposed by using a Kerr nonlinear parametric oscillator (KPO), which is based on a two-photon drive and Kerr nonlinearity~\cite{Goto2016, Goto2016a, Puri2017a, Goto2019a}.
In earlier studies, KPOs have first attracted interest due to their nonlinear dynamics~\cite{Milburn1991, Wielinga1993} and then have been studied for a bosonic code~\cite{Cochrane1999}, where a KPO has been used to generate a superposition of coherent states.
Recent studies have proposed that KPOs can be applied to quantum annealing~\cite{Goto2016, Nigg2017, Puri2017, Zhao2018, Goto2018a, Kewming2020, Onodera2020, Goto2020b, Kanao2021, Goto2021b} and universal quantum computation~\cite{Goto2016a, Puri2017a}.
Furthermore, a set of gate operations for KPOs has been found such that a type of error is suppressed~\cite{Puri2020} and has been developed toward fault-tolerant quantum computation~\cite{Xu2022, Putterman2022}.
For such quantum computation using coherent states, scalable architectures have been studied~\cite{Darmawan2021a, Chamberland2022}.
(Kerr nonlinear oscillators without parametric drive have also been studied for, e.g., the generation of nonclassical states, such as superpositions of coherent states~\cite{Milburn1986, Yurke1986, Miranowicz1990, Greiner2002, Kirchmair2013}, amplitude-squeezed states~\cite{Kitagawa1986}, and finite-dimensional states~\cite{Leonski1994, Imamoglu1997, Leonski2011}.)

A KPO can be implemented by using a superconducting circuit similar to a Josephson parametric oscillator (JPO)~\cite{Yamamoto2008, Lin2014} and has recently been realized in the following experiments.
A superposition of coherent states has first been observed~\cite{Wang2019} and then single-qubit gate operations have been demonstrated~\cite{Grimm2020}.
Also, a crossover from a JPO to a KPO has been observed~\cite{Yamaji2022}.
Related theoretical analyses have been reported~\cite{Masuda2021, Masuda2021b}.

A KPO qubit consists of two coherent states with opposite phases and sufficiently large amplitudes, $|\!\pm\!\alpha\rangle$, which are used as logical basis states $|\tilde{0}\rangle$ and $|\tilde{1}\rangle$.
For these states, the following universal gate set can be implemented: $X$ rotation ($R_x$ gate), $Z$ rotation ($R_z$ gate), and $ZZ$ rotation ($R_{zz}$ gate)~\cite{Mirrahimi2014, Goto2016a, Puri2017a, Masuda2022, Chono2022}.
The $R_z$ and $R_{zz}$ gates can be realized relatively easily by an external driving field with a frequency equal to the resonance frequency of the KPO (single-photon drive) and linear coupling between two KPOs, respectively.

In contrast, an $R_x$ gate, which exchanges populations between the two coherent states, is harder to perform because these states are stabilized by an effective double-well potential due to a parametric drive~\cite{Goto2016}.
The following three methods have been proposed for implementing an $R_x$ gate.
The first is by controlling a detuning frequency of the KPO~\cite{Goto2016a, Puri2017a}.
The second is by turning off the parametric drive for a certain time, utilizing time evolution by Kerr nonlinearity~\cite{Kirchmair2013}, which has been demonstrated experimentally~\cite{Grimm2020}.
The third is by rotating the phase of the parametric drive, which is related to a controlled-NOT gate that preserves the error bias~\cite{Puri2020}.
The first method can execute an ${R_x(\theta)}$ gate with an arbitrary angle $\theta$, while the latter two implement only ${R_x(\pi/2)}$ or ${R_x(\pi)}$ gates.
Although the $R_x(\pi/2)$ gate combined with arbitrary $R_z$ and $R_{zz}$ gates is sufficient for a universal gate set~\cite{Nielsen2000}, the continuous $R_x$ gate will be useful for noisy intermediate-scale quantum (NISQ) algorithms such as variational quantum algorithms (VQAs)~\cite{Cerezo2020, Endo2021}.
However, the continuous $R_x$ gate has not been realized experimentally and thus a simple method for its implementation is needed.

In this paper, we propose an alternative method for a continuous $R_x$ gate by intentionally exciting the KPO outside the qubit space spanned by the two stable coherent states.
Such effective excited states in a rotating frame have not been utilized so far.
This method can realize a continuous $R_x$ gate by only adding a single-photon or two-photon driving field.
We show numerically that the $R_x$ gate can be performed with a high fidelity by utilizing parity selectivity of the excitation.
Using this method, we obtain a faster $R_x$ gate when we use effective excited states rather higher than those nearest to the qubit space.
The numerical results indicate that the two-photon drive gives better performance than the single-photon one.
The proposed method can offer a simple implementation of a continuous $R_x$ gate for a KPO qubit in a superconducting circuit, which will enable flexible designs of, e.g., VQAs.

The paper is organized as follows.
In Sec.~\ref{sec_method}, we introduce the proposed method and in Sec.~\ref{sec_result}, we present the numerical-simulation results.
$R_x$ gates with $|\theta|=\pi/2$ are shown in detail in Sec.~\ref{sec_HighF} and continuous $R_x$ gates are investigated in Sec.~\ref{sec_cont}.
In Sec.~\ref{sec_fast}, we evaluate $R_x$ gates that are faster than the gates in the previous sections.
Finally, in Sec.~\ref{sec_loss}, the effect of single-photon loss is reported.
The paper is concluded with a summary and a brief outlook in Sec.~\ref{sec_summary}.

\section{$R_x$ gate using excited states of KPO}\label{sec_method}
We first describe a Hamiltonian of a KPO and its eigenstates and then propose a method for an $R_x$ gate utilizing effective excited states.
The Hamiltonian is given by~\cite{Goto2016, Puri2017a, Wielinga1993}
\begin{eqnarray}
	\frac{H_{\rm KPO}}{\hbar}&=&-\frac{K}{2}a^{\dagger2}a^2+\frac{p_0}{2}\!\left(a^2+a^{\dagger2}\right),\label{eq_HKPO}
\end{eqnarray}
in a frame rotating at the resonance frequency, $\omega_0$, of the KPO and within the rotating-wave approximation.
Here, $a$ and $a^\dagger$ are, respectively, the annihilation and creation operators for the KPO and $\hbar, K$, and $p_0$ are the reduced Planck constant, the Kerr coefficient, and the amplitude of the parametric drive, respectively.
We choose ${K\!>\!0}$ as in KPOs with superconducting circuits~\cite{Wang2019, Grimm2020, Yamaji2022}.
In the following numerical calculations, we represent operators and states in the photon-number basis with a largest photon number of 30.

\subsection{Effective ground and excited states of KPO}\label{sec_excite}
\begin{figure}
	\includegraphics[width=8cm]{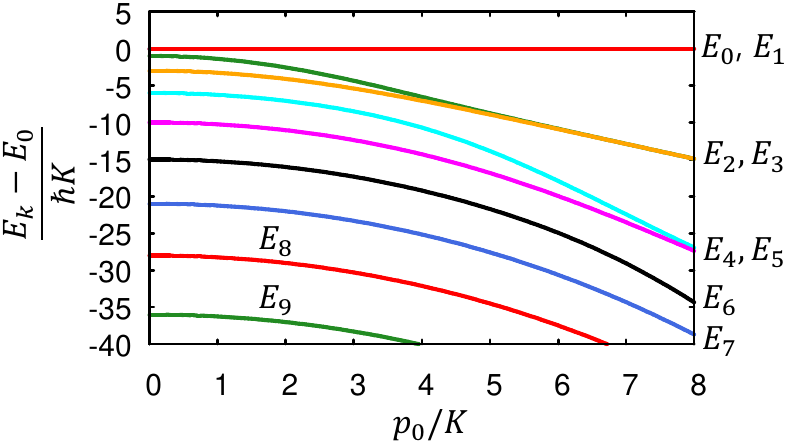}%
	\caption{The eigenvalues of $H_{\rm KPO}$ as functions of $p_0/K$.
		$E_0$ and $E_1$ are degenerate.
		\label{fig_E_p}}
\end{figure}

Figure~\ref{fig_E_p} shows the eigenvalues of $H_{\rm KPO}$ as functions of ${p_0/K}$.
Although $E_0$ and $E_1$ are the degenerate highest eigenvalues in the rotating frame, the corresponding eigenstates can be regarded as effective ground states, because at ${p_0/K=0}$ these states are the eigenstates with the smallest photon numbers, that is, the vacuum and single-photon states, respectively~\cite{Goto2016}.
These states are the ``effective'' ground states because for ${p_0/K\neq0}$, the Hamiltonian in the original laboratory frame depends on time owing to the parametric drive.
In the following, we refer to the states $|E_0\rangle$ and $|E_1\rangle$ as the ground states.

The eigenstates of $H_{\rm KPO}$ can be given by parity eigenstates, because $H_{\rm KPO}$ commutes with the parity operator $e^{i\pi a^\dagger a}$.
The degenerate ground states can be expressed as~\cite{Wielinga1993, Goto2016, Goto2016a, Puri2017a}
\begin{eqnarray}
	|E_0\rangle&=&N_+(|\alpha\rangle+|\!-\!\alpha\rangle),\\
	|E_1\rangle&=&N_-(|\alpha\rangle-|\!-\!\alpha\rangle),
\end{eqnarray}
which have even and odd parities, respectively.
The $|\!\pm\!\alpha\rangle$ are coherent states with $\alpha\!=\!\sqrt{p_0/K}$, satisfying $a|\!\pm\!\alpha\rangle\!=\!\pm\alpha|\!\pm\!\alpha\rangle$, and $N_{\pm}\!=\!1/\sqrt{2\left(1\pm e^{-2\alpha^2}\right)}$ are normalization factors.
We define the computational basis by
\begin{eqnarray}
	|\tilde{0}\rangle&=&\frac{1}{\sqrt{2}}(|E_0\rangle+|E_1\rangle),\\
	|\tilde{1}\rangle&=&\frac{1}{\sqrt{2}}(|E_0\rangle-|E_1\rangle),
\end{eqnarray}
which can be approximated by $|\!\pm\!\alpha\rangle$, respectively, for sufficiently large $\alpha$~\cite{Goto2016a, Puri2017a, Puri2020}.
These basis states are exactly orthogonal, namely, $\langle\tilde{0}|\tilde{1}\rangle\!=\!0$, while $|\!\pm\!\alpha\rangle$ are approximately orthogonal with an exponentially small error of $|\langle\alpha|\!-\!\alpha\rangle|^2\!=\!e^{-4\alpha^2}\!=\!9.2\times10^{-6}$ even for the smallest $p_0/K\!=\!2.9$ in this study.

The rest of the eigenstates can be considered as effective excited states in the rotating frame, which we refer to as the excited states in the following.
These excited states can be detected spectroscopically~\cite{Yamaji2022, Masuda2021b}.
The parity of an excited state $|E_k\rangle$ is the same as that of the integer $k$, because $k$ coincides with the photon number of the eigenstate at $p_0/K\!=\!0$ and the parametric drive preserves the parity.

\subsection{Proposed method}\label{sec_MethodGate}
\begin{figure*}
	\includegraphics[width=13cm]{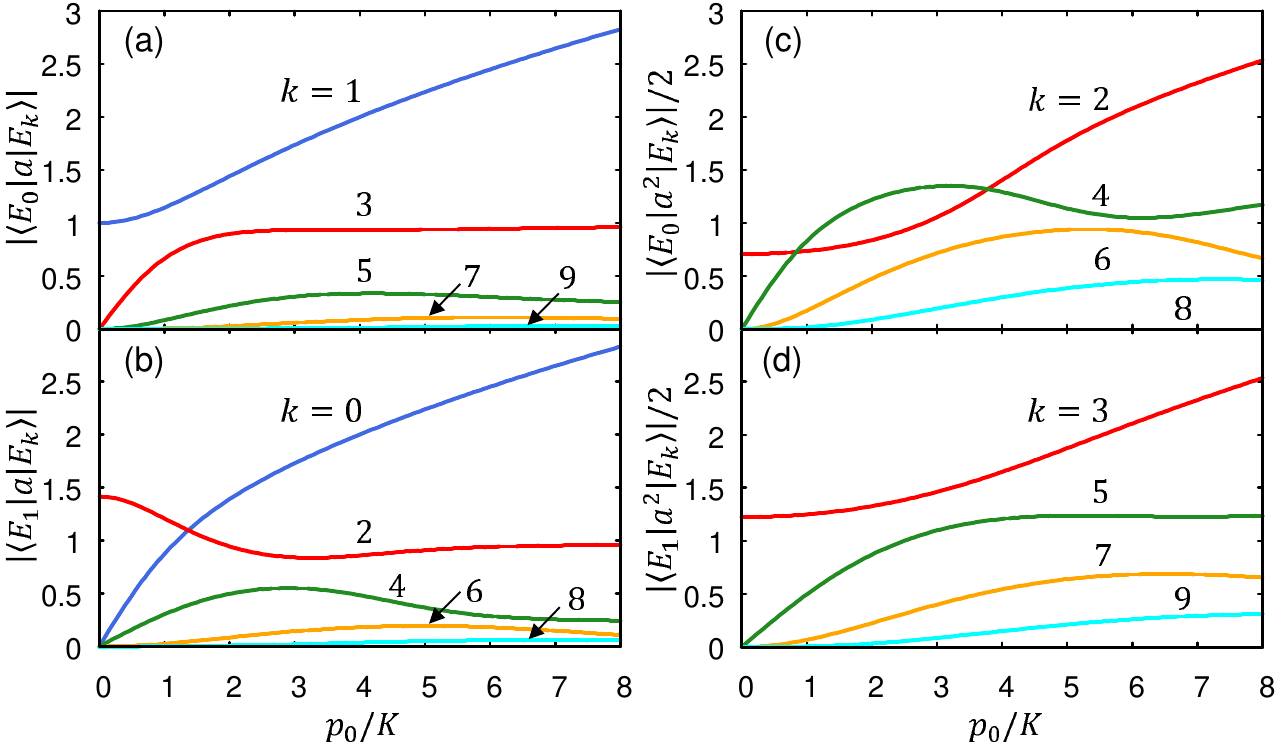}%
	\caption{Matrix elements between ground states and eigenstates of $H_{\rm KPO}$ as functions of $p_0/K$: (a), (b) single-photon and (c), (d) two-photon drives; 	(a), (c) $|E_0\rangle$ and (b), (d) $|E_1\rangle$.
		\label{fig_MatEle_p}}
\end{figure*}
To utilize the excited states for an $R_x$ gate, we add a single-photon or two-photon driving term $H_d$ as
\begin{eqnarray}
	H&=&H_{\rm KPO}+H_d,\label{eq_Ham}\\
	\frac{H_d}{\hbar}&=&Ae^{-i\omega_dt}+A^\dagger e^{i\omega_dt},\label{eq_HamD}
\end{eqnarray}
where $t$, $\omega_d$, and $A$ are, respectively, the time, the driving frequency, and the driving operator given by
\begin{eqnarray}
	A&=&p_d(t)a,\hspace{1em}\text{for the single-photon drive},\\
	A&=&p_d(t)\frac{a^2}{2},\hspace{0.3em}\text{for the two-photon drive},
\end{eqnarray}
with a time-dependent amplitude $p_d(t)$.
These single-photon and two-photon drives can be realized experimentally by applying an external driving field with the frequency ${\omega_0-\omega_d}$ and by adding a two-photon driving field with the frequency $2\omega_0-\omega_d$, respectively.
These fields can be fed through lines used for $R_z$ gates and the parametric drive~\cite{Grimm2020, Yamaji2022}.

$H_d$ can induce parity-selective transitions as follows.
The single-photon drive leads to transitions between states with different parities because the parity of $H_d$ is odd.
For the two-photon drive, $H_d$ has an even parity and induces only parity-conserving transitions.

Figure~\ref{fig_MatEle_p} shows matrix elements of $a$ and $a^2/2$ connected to the ground states, as functions of $p_0/K$.
Note that these matrix elements are finite even for higher excited states such as $k\!=\!4$, $5$, and $6$, which are essential for our proposed method.
For the single-photon drive [Figs.~\ref{fig_MatEle_p}(a) and~\ref{fig_MatEle_p}(b)], large matrix elements are present between the ground states and are approximated by
\begin{eqnarray}
	\langle E_0|a|E_1\rangle\simeq\langle E_1|a|E_0\rangle\simeq\sqrt{\frac{p_0}{K}},\label{eq_MatA}
\end{eqnarray}
for large $p_0/K$.
In contrast, for the two-photon drive, such matrix elements are exactly zero, $\langle E_0|a^2|E_1\rangle\!=\!\langle E_1|a^2|E_0\rangle\!=\!0$, because $a^2$ conserves the parity.
Thus, by tuning $\omega_d$, the two-photon drive can resonantly transfer the population of only one of $|E_0\rangle$ and $|E_1\rangle$ to an excited state, keeping the other ground state almost unchanged.
For the single-photon drive, such an ideal selective transition is prevented by the coupling between the ground states due to the large matrix elements of Eq.~(\ref{eq_MatA}).

We explain the proposed $R_x$ gate first by the ideal selective transitions and then by more realistic transitions.
$R_x(\theta)$, with a rotation angle $\theta$, is expressed in $\{|\tilde{0}\rangle, |\tilde{1}\rangle\}$ as
\begin{eqnarray}
	R_x(\theta)=&\left(\!\begin{array}{cc}
		\displaystyle\cos\!\frac{\theta}{2}\!&\!\displaystyle-i\!\sin\!\frac{\theta}{2}\\
		\\[-2ex]
		\displaystyle\!-i\!\sin\!\frac{\theta}{2}\!&\!\displaystyle\cos\!\frac{\theta}{2}
	\end{array}\!\right).
\end{eqnarray}

Ideally, an $R_x$ gate can be designed as follows.
An initial state is written in $\left\{|E_0\rangle, |E_1\rangle\right\}$ as
\begin{eqnarray}
	|\psi_i\rangle=c_0|\tilde{0}\rangle+c_1|\tilde{1}\rangle=c_+|E_0\rangle+c_-|E_1\rangle,
\end{eqnarray}
where $c_\pm\!=\!\left(c_0\pm c_1\right)/\sqrt{2}$.
We transfer a part of the population of one of the ground states, e.g., $|E_0\rangle$, to an excited state and then return it.
As a result, $|E_0\rangle$ acquires a phase of an angle $\theta_0$ relative to $|E_1\rangle$, generating a final state of
\begin{eqnarray}
	|\psi_f\rangle\!=\!c_+e^{-i\theta_0}\!|E_0\rangle\!+\!c_-|E_1\rangle\!=\!e^{-i\theta_0/2}\!R_x(\theta_0)|\psi_i\rangle.\label{eq_Fin0}
\end{eqnarray}
Hence $R_x(\theta_0)$ is performed (the factor $e^{-i\theta_0/2}$ can be neglected).
Similarly, when $|E_1\rangle$ is selectively coupled with an excited state, $R_x(-\theta_1)$ is realized:
\begin{eqnarray}
	|\psi_f\rangle\!=\!c_+|E_0\rangle\!+\!c_-e^{-i\theta_1}\!|E_1\rangle\!=\!e^{-i\theta_1/2}\!R_x(-\theta_1)|\psi_i\rangle.\label{eq_Fin1}
\end{eqnarray}
By assuming that the driving operator $A$ varies adiabatically, the angle $\theta_g$ with $g\!=\!0$ or $1$ can be expressed as
\begin{eqnarray}
	\theta_g&=&\mathrm{sgn}\!\left(\delta_{ge}\right)\!\int^{T}_0\!dt\!\left(\sqrt{\delta_{ge}^2\!+\!\langle E_g|A|E_e\rangle^2}\!-\!\left|\delta_{ge}\right|\right),\label{eq_theta}\\
	\delta_{ge}&=&\frac{1}{2}\left(\frac{E_g-E_e}{\hbar}-\omega_d\right),\label{eq_delta}
\end{eqnarray}
where $T$ is the gate time and $e$ denotes the utilized excited state (for details, see Appendix~\ref{sec_theta}).
Equation~(\ref{eq_theta}) clarifies that $\theta_g$ originates from a nonzero $\langle E_g|A|E_e\rangle$.
The sign of $\theta_g$ is determined by that of $\delta_{ge}$ because the integrand is positive.

More realistically, we can perform an $R_x$ gate by choosing $\omega_d$ near two excited states.
Populations of $|E_0\rangle$ and $|E_1\rangle$ are separately transferred to these excited states owing to the parity selectivity (when the coupling between the ground states can be neglected).
Then, both $|E_0\rangle$ and $|E_1\rangle$ obtain phases, giving $R_x(\theta_0-\theta_1)$:
\begin{eqnarray}
	|\psi_f\rangle&=&c_+e^{-i\theta_0}|E_0\rangle\!+\!c_-e^{-i\theta_1}|E_1\rangle\\
	&=&e^{-i(\theta_0+\theta_1)/2}R_x(\theta_0-\theta_1)|\psi_i\rangle.\label{eq_Fin01}
\end{eqnarray}
Although Eqs.~(\ref{eq_theta}) and (\ref{eq_Fin01}) provide a picture of the proposed method, we estimate the rotation angle $\theta$ by another method in the numerical simulations in Sec.~\ref{sec_result}, because the precise value of $\theta$ can be affected by transitions to other states, as can be seen shortly.

\begin{figure*}
	\includegraphics[width=14cm]{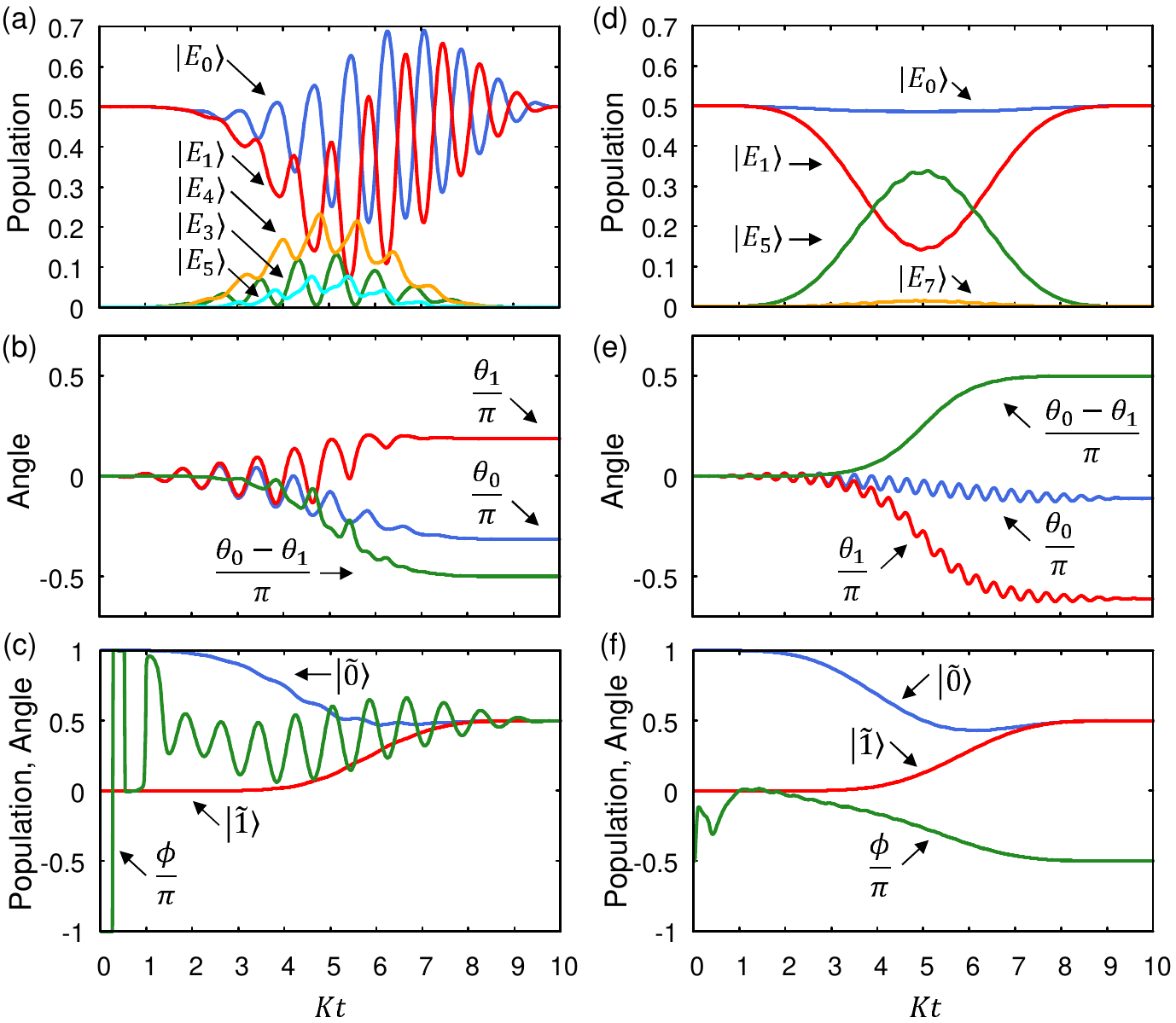}%
	\caption{The time evolutions during $R_x(\theta)$ gates with $|\theta|\!=\!\pi/2$.
		(a)--(c) Single-photon and (d)--(f) two-photon drives.
		(a), (d) Populations in $\left\{|E_k\rangle\right\}$.
		The largest five and four populations, respectively, are plotted, whereas the others are negligibly small.
		(b), (e) The angles $\theta_k\!=\!-\arg\left(\langle E_k|\psi\rangle\right)$.
		(c), (f) The populations in $\left\{|\tilde{0}\rangle, |\tilde{1}\rangle\right\}$, and the relative angles $\phi\!=\!\arg\left(\langle\tilde{1}|\psi\rangle/\langle\tilde{0}|\psi\rangle\right)$.
		$\left(p_0/K, \omega_d/K\right)\!=\!(2.9, 7.79)$ and $(4.7, 16.55)$ for the single-photon and two-photon drives, respectively.
		Pulses in Appendix~\ref{sec_pulse} (Fig.~\ref{fig_pulse_t}) are used.
		\label{fig_state_t}}
\end{figure*}

\section{Numerical simulations}\label{sec_result}
Here, we present our numerical-simulation results of the proposed method.
The gate time $T$ in Secs.~\ref{sec_HighF} and \ref{sec_cont} is $KT\!=\!10$, which is the same as in the previous study~\cite{Goto2016a} (the shortening of $T$ is discussed in Secs.~\ref{sec_fast} and \ref{sec_loss}).
We use a pulse-shaped $p_d(t)$ characterized by a maximum value of $p_{d1}$ and a rise time $\tau$ (for details, see Appendix~\ref{sec_pulse}).

We set the initial state to $|\psi_i\rangle\!=\!|\tilde{0}\rangle$ in order to demonstrate that a part of the population is transferred from $|\tilde{0}\rangle$ to $|\tilde{1}\rangle$.
In Secs.~\ref{sec_HighF}-\ref{sec_fast}, we calculate the time evolution of a state $|\psi\rangle$ by solving the Schr\"{o}dinger equation:
\begin{eqnarray}
	i\hbar\frac{d}{dt}|\psi\rangle&=&H|\psi\rangle.\label{eq_sch}
\end{eqnarray}
From a final state $|\psi_f\rangle$, the fidelity of an $R_x(\theta)$ gate is calculated by 
\begin{eqnarray}
	F=\left|\langle\psi_f|R_x(\theta)|\psi_i\rangle\right|^2
\end{eqnarray}
and the error is defined by $1-F$.
The rotation angle $\theta$ corresponding to parameters $\omega_d$ and $p_d(t)$ is defined by the value of $\theta$ that maximizes $F$~\cite{Goto2016a}, instead of predetermining $\theta$ with Eqs.~(\ref{eq_theta}) and (\ref{eq_Fin01}), as mentioned above.
Once such a correspondence is known, then a desired $R_x(\theta)$ gate can be executed by using the corresponding $\omega_d$ and $p_d(t)$.
We set a criterion for high fidelity by $1-F\!<\!10^{-3}$ and optimize $p_0/K$ and $K\tau$ so that $R_x(\theta)$ gates with $1-F\!<\!10^{-3}$ and $|\theta|\!\geq\!\pi/2$ are achieved in larger areas of $\left(\omega_d/K, p_{d1}/K\right)$.
By these parameters, high-fidelity $R_x(\theta)$ gates with $|\theta|\!<\!\pi/2$ are also obtained, as shown in Sec.~\ref{sec_cont}.

\subsection{$R_x(\theta)$ gates with $|\theta|\!=\!\pi/2$}\label{sec_HighF}
First, we show high-fidelity $R_x(\theta)$ gates with $|\theta|\!=\!\pi/2$, which are sufficient for a universal gate set, as described in Sec.~\ref{sec_intro}.
By the single-photon drive, we obtain an $R_x(-\pi/2)$ with an error of $1-F\!=\!5.1\times10^{-4}$, despite the imperfect selectivity in transitions.
Figure~\ref{fig_state_t}(a) shows that, although the populations oscillate between $|E_0\rangle$ and $|E_1\rangle$ owing to the large matrix elements between these states [Eq.~(\ref{eq_MatA})], the populations almost return to the initial values at the final time.
Also, because the chosen $\omega_d/K$ is close to both $\xi_4$ and $\xi_3$ with 
\begin{eqnarray}
	\xi_k=\frac{E_0-E_k}{\hbar K},\label{eq_xi}
\end{eqnarray}
two excited states $|E_4\rangle$ and $|E_3\rangle$ are largely populated.
As a result, both $|E_0\rangle$ and $|E_1\rangle$ acquire phases with $\theta_g\!=\!-\arg\left(\langle E_g|\psi\rangle\right)$.
These phases then contribute constructively to the total rotation angle $-\pi/2$, as can be seen in Fig.~\ref{fig_state_t}(b), where the signs of $\theta_g$ are consistent with Eq.~(\ref{eq_theta}).
Figure~\ref{fig_state_t}(c) shows that this operation leads to a desired population transfer from $|\tilde{0}\rangle$ to $|\tilde{1}\rangle$, as well as a correct rotation with a relative angle $\phi\!=\!\arg\left(\langle\tilde{1}|\psi\rangle/\langle\tilde{0}|\psi\rangle\right)\!=\!\pi/2$.

For the single-photon drive, the value of $p_0/K\!=\!2.9$ optimized as above is rather small, because the small $p_0/K$ reduces the matrix elements between $|E_0\rangle$ and $|E_1\rangle$ and hence reduces the population oscillation between these states.
In addition, $p_0/K\!=\!2.9$ is optimal in that $\langle E_1|a|E_4\rangle$ is the largest near this value [Fig.~\ref{fig_MatEle_p}(b)], which makes the transition easier.

By the two-photon drive, we obtain a high-fidelity $R_x(\pi/2)$ ($1-F\!=\!5.4\times10^{-4}$).
The time evolution is close to that designed in Sec.~\ref{sec_MethodGate}:
$|E_5\rangle$ is populated almost selectively [Fig.~\ref{fig_state_t}(d)] and the phase of $|E_1\rangle$ dominantly rotates [Fig.~\ref{fig_state_t}(e)], leading to the desired operation [Fig.~\ref{fig_state_t}(f)].

For the two-photon drive, the optimal $p_0/K\!=\!4.7$ is larger, first because no direct transition occurs between $|E_0\rangle$ and $|E_1\rangle$, in contrast to the single-photon drive.
Second, the larger $p_0/K$ widens the separations between the eigenvalues of the lower excited states ($E_2$ and $E_3$) and the higher ones ($E_4$, $E_5$, and $E_6$), as can be seen in Fig.~\ref{fig_E_p}.
These larger separations improve the selectivity of the transition.
Also, the larger separations allow us to use larger $\delta_{ge}$ [Eq.~(\ref{eq_delta})], which can suppress nonadiabatic errors~\cite{Martinis2014}.
When $p_0/K$ is too large, however, the excited states become degenerate in pairs~\cite{Wang2019, Puri2019, Zhang2017}, as in Fig.~\ref{fig_E_p}.
These degenerate states are localized near $\pm\sqrt{p_0/K}$ in the quadrature phase space owing to a double-well potential by the parametric drive~\cite{Wang2019, Puri2019, Zhang2017}.
In other words, the states with large separations of the eigenvalues are the ones extended over the double wells, which we utilize in this method.

Here, the errors are mostly due to leakage, that is, residual populations in excited states.
For KPOs, a method to correct the leakage has been proposed, which applies an artificial two-photon loss~\cite{Puri2019, Puri2020, Xu2022}.
In this paper, for simplicity, we regard the leakage as an error.

\subsection{Continuous $R_x(\theta)$ gates}\label{sec_cont}
\begin{figure}
	\includegraphics[width=8cm]{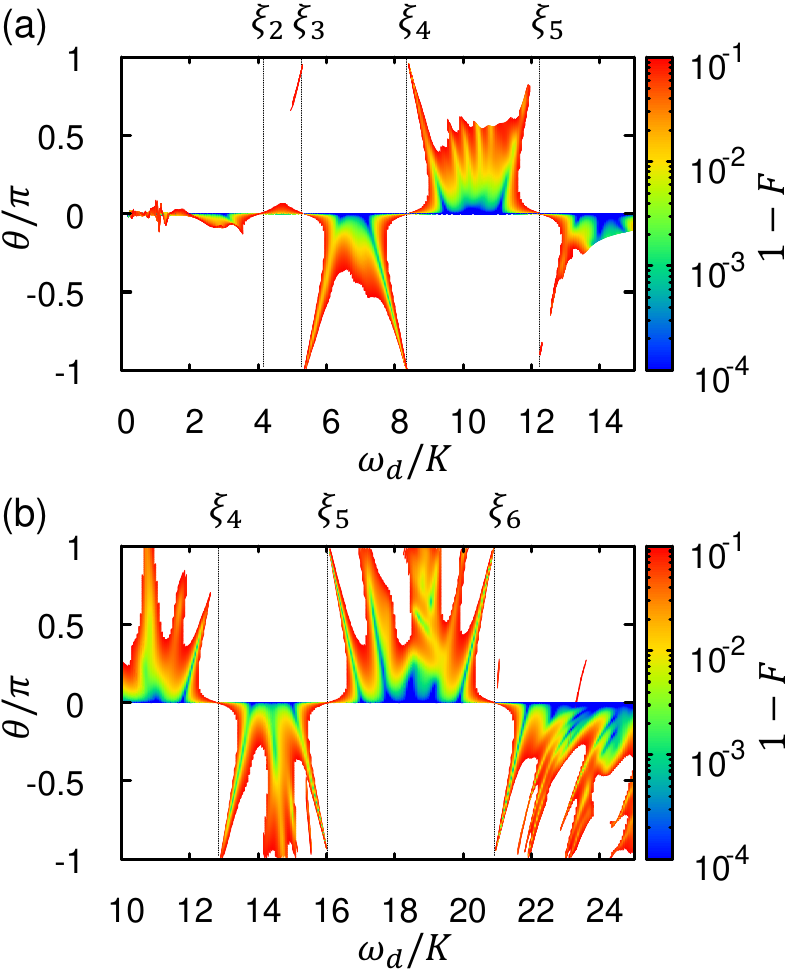}%
	\caption{The rotation angles of $R_x(\theta)$ gates that are obtained with high fidelities as functions of $\omega_d/K$: (a) single-photon and (b) two-photon drives.
		The colors indicate $1-F$. 
		Much higher fidelities $1-F\!<\!10^{-4}$ are plotted using the same color as $1-F\!=\!10^{-4}$ (blue).
		$KT=10$, and $p_{d1}/K$ is swept from $0$ to $2$.
		$\left(p_0/K, K\tau\right)\!=\!(2.9, 3.9)$ and $(4.7, 2.4)$ for the single-photon and two-photon drives, respectively.
		\label{fig_ang_od_drv}}
\end{figure}
Next, we present the results of $R_x(\theta)$ gates with arbitrary rotation angles $\theta$ by changing $\omega_d$ and $p_{d1}$.
Figure~\ref{fig_ang_od_drv} shows the values of $\theta$ that are obtained with high fidelities, as functions of $\omega_d/K$.
The vertical dashed lines indicate $\xi_k$ in Eq.~(\ref{eq_xi}).
For the single-photon drive, continuous rotations between $\theta\!=\!0$ and $-\pi/2$ are possible, with high fidelities of $1-F\!<\!10^{-3}$ by $\omega_d/K$ around $\xi_4$.
For the two-photon drive, such high-fidelity continuous $R_x(\theta)$ gates are made possible by $\omega_d/K$ around $\xi_5$ and $\xi_6$, where $\theta$ can be varied from $0$ to $\pi$, and from $-\pi/2$ to $\pi/2$, respectively.
The signs of $\theta$ can again be understood from Eqs.~(\ref{eq_theta}) and (\ref{eq_Fin01}).

For both the drives, the high-fidelity $R_x(\theta)$ gates with large $|\theta|$ are obtained by $\omega_d/K$ near the higher excited states ($\xi_4, \xi_5$, and $\xi_6$), rather than the lower ones ($\xi_2$ and $\xi_3$), because the separations of the eigenvalues are larger between the higher excited states as mentioned above.
This feature is robust against changes in $p_0/K$ in a certain range.
In the comparison of the two drives shown in Fig.~\ref{fig_ang_od_drv}, the high-fidelity area is larger for the two-photon drive.

\subsection{Faster $R_x$ gates}\label{sec_fast}
Faster gate operations are favorable in suppressing the effect of photon loss.
Here, we evaluate the shortest gate time in the proposed method, keeping the high fidelities of $1-F\!<\!10^{-3}$.
As far as we can search, by the single-photon drive, $R_x(-\pi/2)$ within $KT\!=\!9.1$ is obtained with $p_0/K\!=\!2.9$, $\omega_d/K\!=\!7.78$, $p_{d1}/K\!=\!0.848$, and $K\tau\!=\!2.3$.
By the two-photon drive, $R_x(-\pi/2)$ can be accelerated further to $KT\!=\!6.4$, with $p_0/K\!=\!4.2$, $\omega_d/K\!=\!20.51$, $p_{d1}/K\!=\!0.732$, and $K\tau\!=\!1$.
At this value of $\omega_d/K$, $|E_6\rangle$ is populated, instead of $|E_5\rangle$ used in Sec.~\ref{sec_HighF}.
The faster $R_x$ gate is obtained by the two-photon drive, again because of the absence of the direct transition between $|E_0\rangle$ and $|E_1\rangle$ and the larger separations of the eigenvalues.

\subsection{Effect of single-photon loss}\label{sec_loss}
\begin{figure}
	\includegraphics[width=7cm]{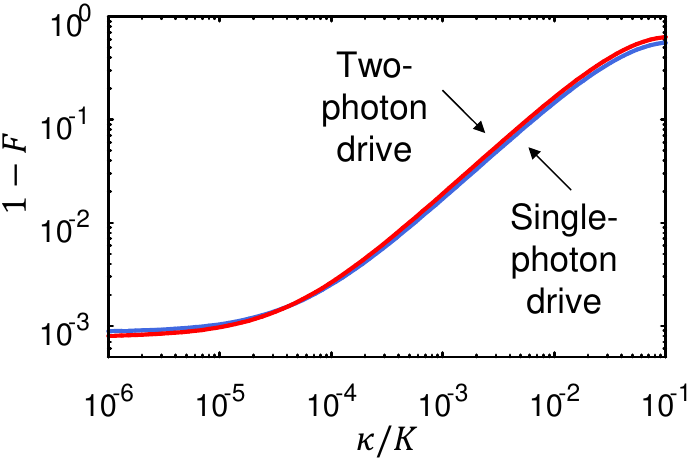}%
	\caption{The errors of the $R_x$ gates as functions of the loss rate $\kappa/K$.
		The parameters are the same as in Sec.~\ref{sec_fast}.
		\label{fig_err_kap_drv}}
\end{figure}

Finally, we study the effect of single-photon loss by solving the master equation for a density operator $\rho$,
\begin{eqnarray}
	\frac{d\rho}{dt}&=&-\frac{i}{\hbar}\left[H,\rho\right]+\frac{\kappa}{2}\left(2a\rho a^\dagger-a^\dagger a\rho-\rho a^\dagger a\right),\label{eq_master}
\end{eqnarray}
where $\kappa$ is the loss rate.
(In this work, photon dephasing~\cite{Puri2017a, Liu2004} is not taken into account, as its effect is typically negligible~\cite{Grimm2020, Yamaji2022}.)
We use the parameters that give the shortest gate times for $R_x(-\pi/2)$ in Sec.~\ref{sec_fast}.
Figure~\ref{fig_err_kap_drv} shows the errors $1-F$ of the $R_x$ gates as functions of $\kappa/K$, where the fidelity is calculated from the density operator at the final time, $\rho_f$, by
\begin{eqnarray}
	F=\langle\psi_i|R_x^\dagger(\theta)\rho_fR_x(\theta)|\psi_i\rangle.
\end{eqnarray}
For both of the drives, the errors depend similarly on $\kappa/K$, which can be understood as follows.
The errors in Fig.~\ref{fig_err_kap_drv} can be approximated by $1-F\!=\!\epsilon+\eta\kappa/K$ for small $\kappa/K$, where $\epsilon$ and $\eta$ are constants independent of $\kappa/K$.
$\epsilon$ is mainly due to the leakage, as mentioned in Sec.~\ref{sec_HighF}.
The latter $\eta\kappa/K$ is caused by the single-photon loss and is proportional to $\alpha^2\kappa T$ when the populations of $|E_0\rangle$ and $|E_1\rangle$ are dominant~\cite{Puri2017a}.
Because the parameters here give almost the same values of $\alpha^2T$ for both drives, the errors exhibit the similar dependence on $\kappa/K$ as in Fig.~\ref{fig_err_kap_drv}.

In the experiments on KPOs reported so far, the smallest $\kappa/K$ value has been $7.7\times10^{-4}$~\cite{Grimm2020}.
This corresponds to $1-F\sim10^{-2}$ in Fig.~\ref{fig_err_kap_drv}.

\section{Summary}\label{sec_summary}
We propose a method for a continuous $R_x$ gate for a KPO by using effective excited states in a rotating frame.
We numerically demonstrate that a high-fidelity $R_x$ gate is obtained with parity-selective transitions.
In this method, the use of higher excited states, rather than states adjacent to ground states, leads to faster execution of the $R_x$ gate.
Although both single-photon and two-photon drives can implement the $R_x$ gate, the latter shows better performance owing to the absence of a direct transition between the ground states and the larger separations of the eigenvalues of the excited states.
This continuous $R_x$ gate will hopefully be useful not only for universal quantum computation but also for NISQ applications.

This method can be realized in a superconducting circuit by adding only a driving field.
The performance is expected to be further improved by optimal controls of the drive, such as shortcuts to adiabaticity~\cite{Guery-Odelin2019, Puri2017a, Goto2019b} and gradient-ascent pulse engineering~\cite{Khaneja2005, Puri2017a}.

\begin{acknowledgments}
We thank Y. Matsuzaki, T. Ishikawa, and H. Chono for valuable discussions.
This paper is based on results obtained from a project, Grant No. JPNP16007, commissioned by the New Energy and Industrial Technology Development Organization (NEDO), Japan.
\end{acknowledgments}

\appendix
\section{ANGLE $\theta_g$}\label{sec_theta}
The angle $\theta_g$ in Eq.~(\ref{eq_theta}) is derived from the Schr\"{o}dinger equation for a subsystem spanned by two relevant eigenstates of $H_{\rm KPO}$.
In the interaction picture $|\psi\rangle_I\!=\!e^{iH_{\rm KPO}t/\hbar}|\psi\rangle$, the Schr\"{o}dinger equation [Eq.~(\ref{eq_sch})] is written as 
\begin{eqnarray}
	i\frac{d\psi_k}{dt}=\sum_l\!\left[A_{kl}e^{i2\delta_{kl}t}\!+\!\left(A^\dagger\right)_{kl}e^{-i2\delta_{lk}t}\right]\!\psi_l,\label{eq_SchInt}
\end{eqnarray}
where $\psi_k\!=\!\langle E_k|\psi\rangle_I$, $A_{kl}\!=\!\langle E_k|A|E_l\rangle$ and $\delta_{kl}\!=\!\left[\left(E_k-E_l\right)/\hbar-\omega_d\right]/2$.
$|\delta_{kl}|$ is supposed to be the smallest for $k\!=\!g$ and $l\!=\!e$, where $g$ and $e$ represent ground and excited states, respectively.
Then, by the rotating-wave approximation, rapidly oscillating terms are dropped.
(As described in Sec.~\ref{sec_HighF}, this approximation is valid for the two-photon drive, but is not so good for the single-photon drive owing to large $A_{01}$ and $A_{10}$.)

According to the parity-selection rules, $A_{ge}\!=\!\gamma$ is supposed to be finite for a ground state $g$, while $A_{\bar{g}e}\!=\!0$ for the other ground state $\bar{g}$.
$\gamma$ can be made positive by choosing the phases of $|E_g\rangle$ and $|E_e\rangle$ and $\gamma\!\ge\!0$ is assumed in the following.
In the two relevant states, Eq.~(\ref{eq_SchInt}) is expressed as
\begin{eqnarray}
	i\frac{d}{dt}\left(\begin{array}{c}
		\psi_g\\
		\psi_e
	\end{array}\right)&=&\left(\begin{array}{cc}
		0&\gamma e^{i2\delta t}\\
		\gamma e^{-i2\delta t}&0
	\end{array}\right)\left(\begin{array}{c}
		\psi_g\\
		\psi_e
	\end{array}\right).
\end{eqnarray}
where $\delta\!=\!\delta_{ge}$.
Then, $\tilde{\psi}_g\!=\!e^{-i\delta t}\psi_g$ and $\tilde{\psi}_e\!=\!e^{i\delta t}\psi_e$ yield
\begin{eqnarray}
	i\frac{d}{dt}\left(\begin{array}{c}
		\tilde{\psi}_g\\
		\tilde{\psi}_e
	\end{array}\right)&=&\left(\begin{array}{cc}
		\delta&\gamma\\
		\gamma&-\delta
	\end{array}\right)\left(\begin{array}{c}
		\tilde{\psi}_g\\
		\tilde{\psi}_e
	\end{array}\right).\label{eq_h}
\end{eqnarray}
The eigenvalues of the matrix in Eq.~(\ref{eq_h}) are $\pm\Omega$, with $\Omega\!=\!\sqrt{\delta^2+\gamma^2}$, and the corresponding eigenvectors are
\begin{eqnarray}
	{\bm u}_\pm&=&\frac{1}{\sqrt{2}}\left(\begin{array}{c}
		\sqrt{1\pm\delta/\Omega}\\
		\pm\sqrt{1\mp\delta/\Omega}
	\end{array}\right).
\end{eqnarray}

We set $\gamma\!=\!0$ at the initial and final times, which determines the initial and final states as ${\bm u}_\pm$ for $\delta\!\gtrless\!0$.
When $\gamma$ varies adiabatically, the state is kept in ${\bm u}_\pm$ and a dynamical phase due to $\pm\Omega$ is accumulated.
As a result, $|E_g\rangle$ obtains a phase factor $e^{-i\theta_g}$ relative to $|E_{\bar{g}}\rangle$, leading to Eq.~(\ref{eq_theta}):
\begin{eqnarray}
	\theta_g&=&\mathrm{sgn}(\delta)\int^{T}_0dt\left(\sqrt{\delta^2+\gamma^2}-|\delta|\right).\end{eqnarray}
\vspace{0.5ex}

\section{PULSE SHAPE}\label{sec_pulse}
\begin{figure}[h]
	\includegraphics[width=7cm]{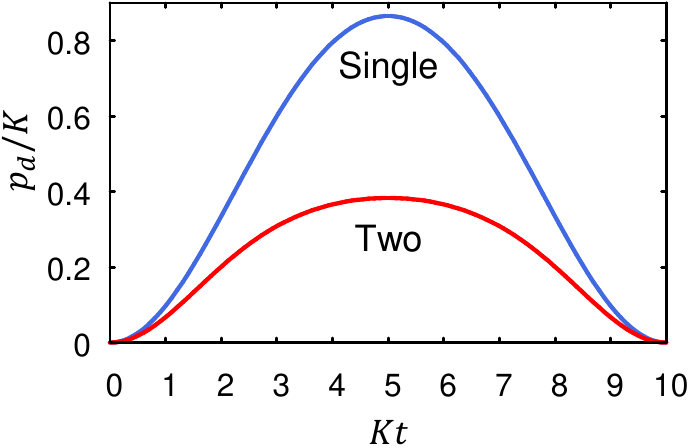}%
	\caption{Pulses with $KT=10$ and $\left(p_{d1}/K, K\tau\right)\!=\!(0.865, 3.9)$ and $(0.383, 2.4)$ for single-photon and two-photon drives, respectively.
		\label{fig_pulse_t}}
\end{figure}

We choose a pulse shape
\begin{eqnarray}
	p_d(t)&=&p_{d1}\left\{\frac{\tanh (t/\tau)\tanh[(T-t)/\tau]}{\tanh^2[T/(2\tau)]}\right\}^2,\label{eq_pulse}
\end{eqnarray}
where $T$ is the gate time and $p_{d1}$ and $\tau$ are the maximum value and the rise time of the pulse, respectively.
By Eq.~(\ref{eq_pulse}), one can generate both a single-peak pulse and a trapezoidlike pulse by changing $\tau$, which is difficult by means of, e.g., a Gaussian pulse shape.
Also, Eq.~(\ref{eq_pulse}) ensures that $p_d(t)$ and $dp_d(t)/dt$ vanish at $t\!=\!0$ and $T$, reducing nonadiabatic errors~\cite{Martinis2014}.
Figure~\ref{fig_pulse_t} shows the pulses used in Sec.~\ref{sec_HighF}.

\end{document}